\documentclass[12pt,amsmath,aps,amssymb,floatfix]{iopart}
\usepackage{bm}

\usepackage{iopams,graphicx}

\begin{document}

\title[Phenomenological Landau analysis of predicted magnetoelectric fluorides]{Phenomenological Landau analysis of predicted magnetoelectric fluorides: KMnFeF$_{6}$ and Ba$_{2}$Ni$_{7}$F$_{18}$}

\author{G. N\'{e}nert\footnote{Gwilherm N\'{e}nert, email address:
gwilherm.nenert@cea.fr.}}

\address{CEA-Grenoble INAC/SPSMS/MDN, 17 rue des martyrs 38054 Grenoble Cedex 9 France\\ (Dated: \today)}

\begin{abstract}\\
Recently, we predicted based on symmetry considerations that
KMnFeF$_{6}$ and Ba$_{2}$Ni$_{7}$F$_{18}$ are likely
magnetoelectric multiferroic materials. In this contribution, we
investigate with Landau theory and crystal structure
considerations the polarization and the linear magnetoelectric
effect in these materials. Based on these two examples, we show
that any magnetoferroelectric will display additional electrical
polarization below its magnetic ordering temperature. This
additional electrical polarization is not related to the linear
magnetoelectric effect. Its magnitude depends on the dielectric
susceptibility.
\end{abstract}

\maketitle

\section{Introduction}

In recent years, the coupling between magnetic and dielectric
properties in transition metal oxides gave rise to a significant
research effort \cite{Fiebig,Eerenstein,Maxim}. This effort is
governed by the emergence of new fundamental physics and potential
technological applications \cite{Eerenstein,Maxim}. This recent
research is mostly concentrated on oxides \cite{nature}. We think
it is very important to search for new materials, especially for
materials different than oxides \cite{nenert,nenert1}. Since
multiferroic and magnetoelectric properties rely on symmetry
considerations, the use of group theory is of prime importance to
look for new materials. The use of symmetry arguments in order to
predict new multiferroic or magnetoelectric materials is not new
\cite{Goshen,Dzialoshinskii,Sergienko1}.

In this contribution, we use the phenomenological Landau theory to
describe the previously predicted magnetoelectric effect in
multiferroic fluorides KMnFeF$_{6}$ and Ba$_{2}$Ni$_{7}$F$_{18}$
\cite{nenert}. These two materials exhibit a pyroelectric symmetry
and thus exhibit a spontaneous electrical polarization at room
temperature. Here we are interested in the behavior of this
electrical polarization at the magnetic ordering temperature. For
this reason, we will consider in the free energy expression only
the lowest degree term of the electrical polarization which is
P$^{2}$. The other coupling terms between the various order
parameters are derived from symmetry considerations. Every product
of the order parameters components belonging to the same
irreducible representation is invariant by time reversal and the
crystal symmetry \cite{Toledano}.

We use the general case of Ba$_{2}$Ni$_{7}$F$_{18}$ (symmetry
$P1$) in order to discuss various contributions to the induced
polarization. In particular we show that any magnetically ordered
pyroelectric materials will display additional electrical
polarization below its magnetic ordering temperature. This
electric polarization is not related to the linear magnetoelectric
effect. We discuss these results in the light of recent reports on
magnetoelectric and multiferroic materials
\cite{Nedlin,Moskvin,Liu,Hu,Zhong}.

\section{Study of KMnFeF$_{6}$}\label{Study of KMnFeF$_{6}$}

The crystal structure of the fluoride KMnFeF$_{6}$ has been
discussed in detail in literature \cite{lacorre2}. This compound
crystallizes in the space group $Pba2$ (n$^{\circ}$32), where the
Mn and Fe ions order on the 4c Wyckoff position of the structure
and occupy statistically the 4b Wyckoff position. KMnFeF$_{6}$
orders ferrimagnetically below T$_{C}$ = 148 K with a ratio
$\frac{\Theta}{T_{C}}$=3. The magnetic unit-cell is identical to
the chemical unit cell and thus $\overrightarrow{k}$ =
$\overrightarrow{0}$. A symmetry analysis by Bertaut's method
gives rise to the results presented in Tables \ref{table1} and
\ref{table2} \cite{lacorre2,Bertaut}.

\begin{table}[htb]
\centering
\begin{tabular}{|{c}|{c}|{c}|{c}|{c}|}
\hline
Modes & x  & y & z & Magnetic space groups\\
\hline
$\Gamma_{1}$     & G$_{x}$  &   A$_{y}$  &   C$_{z}$ & $Pba2$ \\
\hline
$\Gamma_{2}$     & C$_{x}$  &   F$_{y}$  &   G$_{z}$ & $Pba'2'$ \\
\hline
$\Gamma_{3}$     & A$_{x}$  &   G$_{y}$  &   F$_{z}$ & $Pb'a'2$  \\
\hline
$\Gamma_{4}$     & F$_{x}$  &   C$_{y}$  &   A$_{z}$ & $Pb'a2'$  \\
\hline
\end{tabular}
\\
\caption{Irreducible representations for the space group $Pba21'$
associated to \textbf{k}=0 for the Wyckoff position 4c
\cite{lacorre2}.}\label{table1}
\end{table}

\begin{table}[htb]
\centering
\begin{tabular}{|{c}|{c}|{c}|{c}|{c}|}
\hline
Modes & x  & y & z & Magnetic space groups\\
\hline
$\Gamma_{1}$     & -  &   -  &   C$_{z}$ & $Pba2$ \\
\hline
$\Gamma_{2}$     & C$_{x}$  &   F$_{y}$  &   - & $Pba'2'$ \\
\hline
$\Gamma_{3}$     & -  &   -  &   F$_{z}$ & $Pb'a'2$  \\
\hline
$\Gamma_{4}$     & F$_{x}$  &   C$_{y}$  &   - & $Pb'a2'$  \\
\hline
\end{tabular}
\\
\caption{Irreducible representations for the space group $Pba21'$
associated to \textbf{k}=0 for the Wyckoff position 2b
\cite{lacorre2}.}\label{table2}
\end{table}

For the Wyckoff position 4c, the labelling of the various spins is
$\overrightarrow{S_{1}}$ in (x,y,z); $\overrightarrow{S_{2}}$ in
($\overline{x}$,$\overline{y}$,z); $\overrightarrow{S_{3}}$ in
(1/2-x,1/2+y,z) and $\overrightarrow{S_{4}}$ in (1/2+x,1/2-y,z).
The authors defined the following magnetic vectors:

\begin{eqnarray}
\label{eq1}
\eqalign{\overrightarrow{F}=\overrightarrow{S_{1}}+\overrightarrow{S_{2}}+\overrightarrow{S_{3}}+\overrightarrow{S_{4}}\\
\overrightarrow{G}=\overrightarrow{S_{1}}-\overrightarrow{S_{2}}+\overrightarrow{S_{3}}-\overrightarrow{S_{4}}\\
\overrightarrow{C}=\overrightarrow{S_{1}}+\overrightarrow{S_{2}}-\overrightarrow{S_{3}}-\overrightarrow{S_{4}}\\
\overrightarrow{A}=\overrightarrow{S_{1}}-\overrightarrow{S_{2}}-\overrightarrow{S_{3}}+\overrightarrow{S_{4}}}
\end{eqnarray}

For the Wyckoff position 2b, the authors defined two magnetic
vectors
$\overrightarrow{F'}=\overrightarrow{S_{1}}+\overrightarrow{S_{3}}$
and
$\overrightarrow{C'}=\overrightarrow{S_{1}}-\overrightarrow{S_{3}}$
where this time $\overrightarrow{S_{1}}$ is in (0,1/2,z) while
$\overrightarrow{S_{3}}$ is in (1/2,0,z). The decomposition is
given in Tables \ref{table1} and \ref{table2}. The neutron data
show that the best model for the magnetic structure is given by
the $\Gamma_{4}$ mode. The various components of the magnetic
vectors defined in (\ref{eq1}) transform as the reducible
representation. The appropriate way is to consider vectors which
transform as irreducible representation (IR) \cite{Toledano}. In
order to simplify the expression of the free energy and fulfill
the symmetry requirements, we defined the magnetic order parameter
$\eta$ which transform as the $\Gamma_{4}$ IR as observed
experimentally. The total magnetization of the system is defined
by the order parameter $\overrightarrow{M}$. Its components
transform like the components of $\overrightarrow{F}$ (tables
\ref{table1} and \ref{table2}). Thus M$_{x}$, M$_{y}$ and M$_{z}$
transform like the IRs $\Gamma_{4}$, $\Gamma_{2}$ and $\Gamma_{3}$
respectively.

$\eta$ and M$_{x}$ transform both as the $\Gamma_{4}$ IR, thus the
lowest coupling term between $\eta$ and $\overrightarrow{M}$ is
$\eta$ M$_{x}$. The crystal structure of KMnFeF$_{6}$ has an
orthorhombic pyroelectric symmetry. Thus this material exhibits a
spontaneous electrical polarization along the z axis. All the
symmetry elements of the space group \emph{Pba2} let invariant
only the P$_{z}$ component of the electrical polarization. Thus
the lowest coupling term between the magnetic order parameter
$\eta$ and the polarization is $\eta^{2}$P$_{z}$. For the possible
magnetoelectric coupling terms, one has to consider the possible
terms of the type $\eta$MP \cite{magnetoelectrics}. It can be
shown that the electrical polarization components P$_{x}$, P$_{y}$
and P$_{z}$ belong respectively to the IRs $\Gamma_{3}$,
$\Gamma_{4}$ and $\Gamma_{1}$ respectively. Looking at the
possible  $\eta$MP and taking into account the above
considerations, we can write the free energy for KMnFeF$_{6}$ in
the presence of a magnetic field as:

\begin{eqnarray}
\label{eq2}
\eqalign{\Phi&=\Phi_{0}+\frac{a}{2}\eta^{2}+\frac{b}{4}\eta^{4}+\gamma\eta
M_{x}+\sigma\eta^{2}P_{z}+\sum_{i,j}\frac{\kappa_{i,j}}{2}P_{j}^{2}\\
&+\sum_{i,j}\frac{c_{i,j}}{2}M_{j}^{2}+\lambda_{1}\eta
P_{y}M_{z}+\lambda_{2}\eta P_{z}M_{x}-\mathbf{M}.\mathbf{H}}
\end{eqnarray}

We find the partial derivatives of $\Phi$ at the equilibrium
conditions.

\begin{eqnarray}
\label{eq3} \eqalign{\frac{\partial \Phi}{\partial\eta}=a\eta +
b\eta^{3} + \gamma M_{x} + 2\sigma\eta P_{z} +
\lambda_{1}M_{z}P_{y} +
\lambda_{2}M_{x}P_{z}=0\\
\frac{\partial \Phi}{\partial P_{x}}=\kappa_{11}P_{x} = 0\\
\frac{\partial \Phi}{\partial P_{y}}=\kappa_{22}P_{y}+\lambda_{1}\eta M_{z} = 0\\
\frac{\partial \Phi}{\partial P_{z}}=\kappa_{33}P_{z}+\lambda_{2}\eta M_{x}+\sigma\eta^{2}= 0\\
\frac{\partial \Phi}{\partial M_{x}}=c_{11}M_{x}+\lambda_{2}\eta P_{z}+\gamma\eta-H_{x}=0\\
\frac{\partial \Phi}{\partial M_{y}}=c_{22}M_{y}-H_{y}=0\\
\frac{\partial \Phi}{\partial M_{z}}=c_{33}M_{z} + \lambda_{1}\eta
P_{y}-H_{z}=0}
\end{eqnarray}

In order to investigate the linear magnetoelectric coupling, we
express the electrical polarization as function of the magnetic
field. For this purpose, we use the expressions of $\frac{\partial
\Phi}{\partial P_{z}}$ and of $\frac{\partial \Phi}{\partial
P_{y}}$. We find:

\begin{eqnarray}
\label{eq4}
\eqalign{P_{y}=\frac{-\lambda_{1}}{\kappa_{22}}\eta M_{z}\\
P_{z}=\frac{-(\sigma\eta^{2}+\lambda_{2}\eta M_{x})}{\kappa_{33}}}
\end{eqnarray}

Using the results of equations (\ref{eq3}) and (\ref{eq4}), we can
derive the expression for the various components of the
magnetization.

\begin{eqnarray}
\label{eq5}
\eqalign{M_{x}=\frac{\kappa_{33}}{\kappa_{33}c_{11}-\lambda_{2}^{2}\eta^{2}}H_{x}+\frac{\eta(\lambda_{2}\sigma\eta^{2}-\kappa_{33}\gamma)}{\kappa_{33}c_{11}-\lambda_{2}^{2}\eta^{2}}\\
M_{y}=\frac{H_{y}}{c_{22}}\\
M_{z}=\frac{\kappa_{22}}{\kappa_{22}c_{22}-\lambda_{1}^{2}\eta^{2}}H_{z}}
\end{eqnarray}

From equations (\ref{eq4}) and (\ref{eq5}), we can find the
expression for the linear relationship between the induced
polarization and the application of a magnetic field (linear
magnetoelectric effect) for the fluoride KMnFeF$_{6}$.

\begin{eqnarray}
\label{eq6}
\eqalign{P_{x}=0\\
P_{y}=\frac{-\lambda_{1}\eta}{\kappa_{22}c_{33}-\lambda_{1}^{2}\eta^{2}}H_{z}\\
P_{z}=\frac{-\lambda_{2}\eta}{\kappa_{33}c_{11}-\lambda_{2}^{2}\eta^{2}}H_{x}+\frac{(\lambda_{2}\gamma-\sigma
c_{11})\eta^{2}}{\kappa_{33}c_{11}-\lambda_{2}^{2}\eta^{2}}}
\end{eqnarray}

From equation (\ref{eq6}), we find that there are two non-zero
components for the linear magnetoelectric tensor of KMnFeF$_{6}$:
$\alpha_{23}$ and $\alpha_{31}$. $\alpha_{23}$ takes the value
$\frac{-\lambda_{1}\eta}{\kappa_{22}c_{33}-\lambda_{1}^{2}\eta^{2}}$
and $\alpha_{31}$ is equal to
$\frac{-\lambda_{2}\eta}{\kappa_{33}c_{11}-\lambda_{2}^{2}\eta^{2}}$.
We notice that the polarization along x is zero at any temperature
even under the application of a magnetic field. Our results
suggest that the polarization along y is purely induced by the
magnetic field (linear magnetoelectric effect) while the
polarization along z would have an additional component arising at
the magnetic transition temperature (term in $\eta^{2}$ in
equation (\ref{eq6})). We notice that this additional electrical
polarization along z arising below T$_{N}$ is present even in the
absence of a linear magnetoelectric effect ($\lambda_{i}$ = 0). We
stress that our model describes only what is happening below the
magnetic ordering temperature. One should keep in mind that there
is already a spontaneous polarization along z at room temperature
resulting from the pyroelectric symmetry exhibited by
KMnFeF$_{6}$. We notice that the magnetic point group m'm2'
described by the irreducible representation $\Gamma_{4}$ should
exhibit two non-zero components for the magnetoelectric tensor
according to Ref. \cite{ITC}: $\alpha_{23}$ and $\alpha_{32}$
instead of $\alpha_{31}$. This difference can be explained by the
fact that we have used the labelling of the authors of Ref.
\cite{lacorre2}. If we inverse $\overrightarrow{S_{3}}$ and
$\overrightarrow{S_{4}}$, we find that the total magnetization
component M$_{y}$ transforms like $\Gamma_{4}$ instead of
$\Gamma_{2}$ (see tables \ref{table1} and \ref{table2}).
Consequently the invariant term responsible for the linear
magnetoelectric effect is not $\lambda_{2}\eta P_{z}M_{x}$ but
$\lambda_{2}\eta P_{z}M_{y}$ which will give rise to $\alpha_{32}$
instead of $\alpha_{31}$. In addition to this linear
magnetoelectric effect, we notice that a spontaneous magnetization
component along x can be displayed below the N\'{e}el temperature
in the absence of magnetic field (see Eq. (\ref{eq5})).

\section{Study of Ba$_{2}$Ni$_{7}$F$_{18}$}

\subsection{Estimation of the spontaneous polarization
\protect\footnote{We notice that this estimation is not possible
for KMnFeF$_{6}$ since in the reported crystal structure
\cite{lacorre2}, the authors fixed the cations on the atomic
positions determined in the $P4_{2}bc$ symmetry and not in $Pba2$.
Consequently a redetermination of the crystal structure of
KMnFeF$_{6}$ is required to investigate properly the coupling
between dielectric and magnetic properties.}}

To estimate the spontaneous polarization of a pyroelectric system,
one may deduce the possible high-symmetry (or high-temperature)
structure from the low-symmetry structure based on a
pseudosymmetry analysis. This concept gave rise to the prediction
of a large number of displacive ferroelectrics
\cite{ferroelectrics,pseudo}. It has been implemented in the
program PSEUDO (Crystallographic Bilbao Server) \cite{pseudo}.
When the atomic coordinates of a given structure display an
approximate symmetry in addition to the actual space-group
symmetry, the structure can be considered as pseudosymmetric with
respect to a supergroup containing this additional symmetry. The
existence of pseudosymmetry in a crystal structure indicates a
slightly distorted structure of higher symmetry. If the distortion
is small enough, one can expect the crystal to acquire this higher
symmetry at a higher temperature. Using the crystal structure of
Ba$_{2}$Ni$_{7}$F$_{18}$ reported by Lacorre \emph{et al.}
\cite{lacorre4}, we performed a pseudosymmetry search among all
the minimal supergroups of $P1$ (symmetry exhibited by
Ba$_{2}$Ni$_{7}$F$_{18}$). Of the minimal supergroups of $P1$, we
determined $P\overline{1}$ was the only pseudosymmetric minimal
supergroup. We therefore concluded the high-temperature
paraelectric phase of Ba$_{2}$Ni$_{7}$F$_{18}$ is in this space
group. However, this pseudosymmetric minimal supergroup would
produce a maximum atomic displacement of 1.3 \r{A}. The high value
of the atomic displacement towards $P\overline{1}$ suggests that
the paraelectric phase will not be reached before melting or
decomposition. However, the possibility of twinning and domain
switching may exist. The pseudosymmetric minimal supergroup allows
us to estimate the spontaneous polarization in
Ba$_{2}$Ni$_{7}$F$_{18}$. This is can be done using an ionic model
by simplifying the electric charge of each ion with a point charge
\cite{polarization}.

\begin{eqnarray}
\label{polarisation}
P_{s}=\left(\frac{e}{V}\right)\sum_{i}m_{i}Q_{i}\Delta z_{i}
\end{eqnarray}

e is the elementary charge, V the unit-cell volume, m$_{i}$ the
multiplicity of the ion, Q$_{i}$ the ionic charge and
$\Delta$z$_{i}$ the displacement ion along the polar axis. Having
discussed already the hypothetical high temperature paraelectric
phase, we can estimate the displacement $\Delta$z$_{i}$ for each
ion. In that case, since we have the symmetry $P1$, we have to
deal with a spontaneous polarization vector having three non-zero
components: P$_{x}$, P$_{y}$ and P$_{z}$. Using the point charge
model, we estimate that P$_{x}$ = 0.246 $\mu$C/cm$^{2}$; P$_{y}$ =
10.87 $\mu$C/cm$^{2}$ and P$_{z}$ = -26.87 $\mu$C/cm$^{2}$. The
total polarization is thus $\|\overrightarrow{P}\|$ $\cong$ 29
$\mu$C/cm$^{2}$. This is a large spontaneous polarization but this
value is an upper bound. The ionic model used is a crude picture
which does not take into account the covalency of the bonds.
Additionally, the calculation is strongly dependent of the
accuracy of the z coordinate. For BaAl$_{2}$O$_{4}$, the ionic
model gives an overestimation by a factor of six of P$_{s}$
compared to the experimental value \cite{BaAl2O4}. Nevertheless,
the spontaneous polarization in Ba$_{2}$Ni$_{7}$F$_{18}$ remains
high and comparable to known multferroic systems such as
YMnO$_{3}$ (P$_{s}$ = 5.5 $\mu$C/cm$^{2}$) \cite{YMnO3}.

\subsection{Landau phenomenological description of the linear magnetoelectric effect}

In the previous fluorides, the magnetic frustration appeared in
corner-sharing octahedra through the existence of interconnected
triangles of magnetic ions. This leads to a single type of
magnetic exchange interaction. P. Lacorre and coworkers have been
also investigating compounds like Ba$_{2}$Ni$_{3}$F$_{10}$ (n = 9)
and Ba$_{2}$Ni$_{7}$F$_{18}$ (n = 21) which are members of the
Ba$_{6}$M$_{n}$F$_{12+2n}$ family \cite{lacorre4,lacorre3}. In
this family where M = Ni, there are not only corner-sharing
octahedra but also edge-sharing octahedra. Both types of
interaction exist in the Ba$_{2}$Ni$_{3}$F$_{10}$ and
Ba$_{2}$Ni$_{7}$F$_{18}$ compounds. These compounds have been
investigated by means of powder neutron diffraction at room and
low temperatures.

\begin{table}[htb]
\centering
\begin{tabular}{|{c}|{c}|{c}|}
\hline
& h$_{1}$& h$_{1}^{'}$  \\
\hline
 $\Gamma_{1}$     & 1 & -1\\
\hline
\end{tabular}
\\
\caption{Irreducible representation for the space group $P11'$
associated to \textbf{k}=(0, 0, 0).}\label{P11}
\end{table}

In order to keep consistency in the notations between the two
cases, we will call also $\eta$ the magnetic order parameter
describing the magnetic ordering of Ba$_{2}$Ni$_{7}$F$_{18}$.
Since there is only one IR in that case, following similar
arguments that in the case of KMnFeF$_{6}$; all the coupling terms
$\eta$M$_{j}$ are invariant. Since the crystal structure is $P1$,
spontaneous electrical polarization components are non-zero for
all the three directions x, y and z. Consequently, the invariant
terms involving the electrical polarization are
$\sum_{i,j}\frac{\kappa_{i,j}}{2}P_{j}^{2}$ and
$\sum_{i}\sigma_{i}\eta^{2}P_{i}$. In addition to these terms, one
needs to consider the magnetoelectric coupling terms. All the
electrical polarization components transform as the unique
$\Gamma_{1}$ IR. Thus all $\eta P_{i}M_{j}$ are allowed.
Consequently we can write the free energy for this fluoride as:

\begin{eqnarray}
\label{eq7}
\eqalign{\Psi&=\Psi_{0}+\frac{a}{2}\eta^{2}+\frac{b}{4}\eta^{4}+\left(\sigma_{x}P_{x}+\sigma_{y}P_{y}+\sigma_{z}P_{z}\right)\eta^{2}+\sum_{i,j}\frac{\kappa_{i,j}}{2}P_{j}^{2}+\sum_{i,j}\frac{c_{i,j}}{2}M_{j}^{2}\\
&+\left(\gamma_{x}M_{x}+\gamma_{y}M_{y}+\gamma_{z}M_{z}\right)\eta+\left(\lambda_{1}P_{x}M_{x}+\lambda_{2}P_{x}M_{y}+\lambda_{3}P_{x}M_{z}\right)\eta\\
&+\left(\lambda_{4}P_{y}M_{x}+\lambda_{5}P_{y}M_{y}+\lambda_{6}P_{y}M_{z}+\lambda_{7}P_{z}M_{x}\right)\eta\\
&+\left(\lambda_{8}P_{z}M_{y}+\lambda_{9}P_{z}M_{z}\right)\eta-M.H}
\end{eqnarray}

\textbf{M} and \textbf{P} designate respectively the total
magnetization and the electrical polarization of the system.
Calculating the derivatives of the different variables, we can
write the following set of equations:

\begin{eqnarray}
\label{eq8} \eqalign{\frac{\partial \Psi}{\partial\eta}&=a\eta+b\eta^{3}+2\eta\left(\sigma_{x}P_{x}+\sigma_{y}P_{y}+\sigma_{z}P_{z}\right)+\gamma_{x}M_{x}+\gamma_{y}M_{y}+\gamma_{z}M_{z}\\
&+\lambda_{1}P_{x}M_{x}+\lambda_{2}P_{x}M_{y}+\lambda_{3}P_{x}M_{z}+\lambda_{4}P_{y}M_{x}+\lambda_{5}P_{y}M_{y}+\lambda_{6}P_{y}M_{z}\\
&+\lambda_{7}P_{z}M_{x}+\lambda_{8}P_{z}M_{y}+\lambda_{9}P_{z}M_{z}=0\\
\frac{\partial \Psi}{\partial
P_{x}}&=\sigma_{x}\eta^{2}+\kappa_{x}P_{x}+\left(\lambda_{1}M_{x}+\lambda_{2}M_{y}+\lambda_{3}M_{z}\right)\eta
= 0\\
\frac{\partial \Psi}{\partial
P_{y}}&=\sigma_{y}\eta^{2}+\kappa_{y}P_{y}+\left(\lambda_{4}M_{x}+\lambda_{5}M_{y}+\lambda_{6}M_{z}\right)\eta
= 0\\
\frac{\partial \Psi}{\partial
P_{z}}&=\sigma_{z}\eta^{2}+\kappa_{z}P_{z}+\left(\lambda_{7}M_{x}+\lambda_{8}M_{y}+\lambda_{9}M_{z}\right)\eta
= 0\\
\frac{\partial \Psi}{\partial
M_{x}}&=c_{x}M_{x}+\gamma_{x}\eta+\left(\lambda_{1}P_{x}+\lambda_{2}P_{y}+\lambda_{3}P_{z}\right)\eta-H_{x}=0\\
\frac{\partial \Psi}{\partial
M_{y}}&=c_{y}M_{y}+\gamma_{y}\eta+\left(\lambda_{4}P_{x}+\lambda_{5}P_{y}+\lambda_{6}P_{z}\right)\eta-H_{y}=0\\
\frac{\partial \Psi}{\partial
M_{z}}&=c_{z}M_{z}+\gamma_{z}\eta+\left(\lambda_{7}P_{x}+\lambda_{8}P_{y}+\lambda_{9}P_{z}\right)\eta-H_{z}=0}
\end{eqnarray}

Here the dielectric tensor $\kappa_{i,j}$ has 9 non zero terms. To
simplify the notation, we write
$\kappa_{x}$=$\kappa_{11}$+$\kappa_{12}$+$\kappa_{13}$,
$\kappa_{y}$=$\kappa_{21}$+$\kappa_{22}$+$\kappa_{23}$ and
$\kappa_{z}$=$\kappa_{31}$+$\kappa_{32}$+$\kappa_{33}$. We use a
similar notation for c$_{x}$, c$_{y}$ and c$_{z}$.In a similar
manner as in the case of KMnFeF$_{6}$, we can extract from
equation (\ref{eq8}), the expression of M$_{i}$ as function of the
magnetic field, of the magnetic order parameter and of the
polarization. We find the expressions for the various magnetic
components:

\begin{eqnarray}
\label{eq9}
\eqalign{M_{x}=\frac{H_{x}-\eta\left(\gamma_{x}+\lambda_{1}P_{x}+\lambda_{4}P_{y}+\lambda_{7}P_{z}\right)}{c_{x}}\\
M_{y}=\frac{H_{y}-\eta\left(\gamma_{y}+\lambda_{2}P_{x}+\lambda_{5}P_{y}+\lambda_{8}P_{z}\right)}{c_{y}}\\
M_{z}=\frac{H_{z}-\eta\left(\gamma_{z}+\lambda_{3}P_{x}+\lambda_{6}P_{y}+\lambda_{9}P_{z}\right)}{c_{z}}}
\end{eqnarray}

If we replace the various M$_{i}$ components in the expressions of
$\frac{\partial F}{\partial P_{j}}$, we will find, for instance,
P$_{x}$ as function of P$_{y}$ and P$_{z}$. To determine the
expression of P$_{x}$ as function only of the various constants
and of the magnetic order parameter $\eta$, we need to express
P$_{y}$ and P$_{z}$ as function of P$_{x}$. After we can use these
results to solve $\frac{\partial F}{\partial P_{x}} = 0$ as the
only function of P$_{x}$. We have in total six equations with six
variables. This system is solvable exactly. However, the full
expression of the various electrical polarization components is
very lengthy and complicated. Consequently, in order to simplify
the discussion we made a series expansion of P$_{x}$, P$_{y}$ and
P$_{z}$ in terms of $\eta$ at the second order:

\begin{eqnarray}
\label{eq10}
\eqalign{&P_{x}\cong\left(\frac{-\lambda_{1}}{c_{x}\kappa_{x}}H_{x}+\frac{-\lambda_{2}}{c_{y}\kappa_{x}}H_{y}+\frac{-\lambda_{3}}{c_{z}\kappa_{x}}H_{z}\right)\eta\\
&+\left(\frac{-\sigma_{x}}{\kappa_{x}}+\frac{\lambda_{1}\gamma_{x}}{c_{x}\kappa_{x}}+\frac{\lambda_{2}\gamma_{y}}{c_{y}\kappa_{x}}+\frac{\lambda_{3}\gamma_{z}}{c_{z}\kappa_{x}}\right)\eta^{2};\\
&P_{y}\cong\left(\frac{-\lambda_{4}}{c_{x}\kappa_{y}}H_{x}+\frac{-\lambda_{5}}{c_{y}\kappa_{y}}H_{y}+\frac{-\lambda_{6}}{c_{z}\kappa_{y}}H_{z}\right)\eta\\
&+\left(\frac{-\sigma_{y}}{\kappa_{y}}+\frac{\lambda_{4}\gamma_{x}}{c_{x}\kappa_{y}}+\frac{\lambda_{5}\gamma_{y}}{c_{y}\kappa_{y}}+\frac{\lambda_{6}\gamma_{z}}{c_{z}\kappa_{y}}\right)\eta^{2}\\
\mbox{and} \quad
&P_{z}\cong\left(\frac{-\lambda_{7}}{c_{x}\kappa_{z}}H_{x}+\frac{-\lambda_{8}}{c_{y}\kappa_{z}}H_{y}+\frac{-\lambda_{9}}{c_{z}\kappa_{z}}H_{z}\right)\eta\\
&+\left(\frac{-\sigma_{z}}{\kappa_{z}}+\frac{\lambda_{7}\gamma_{x}}{c_{x}\kappa_{z}}+\frac{\lambda_{8}\gamma_{y}}{c_{y}\kappa_{z}}+\frac{\lambda_{9}\gamma_{z}}{c_{z}\kappa_{z}}\right)\eta^{2}}
\end{eqnarray}

From the results of equation (\ref{eq10}), we are able to
determine the magnetoelectric tensor [$\alpha_{i,j}$]. All the
terms of the tensor are non-zero. This is in agreement with the
results of Ref. \cite{ITC}. We find a similar result than in the
case of KMnFeF$_{6}$ where the various component of the tensor are
equal to $\frac{\lambda_{k}}{c_{i}\kappa_{j}}\eta$. This result is
equivalent to the one reported in section \ref{Study of
KMnFeF$_{6}$} in the hypothesis of small value of $\eta$. In a
similar way, there are three non zero spontaneous magnetization
components along the three directions x, y and z (M$_{j}\neq$0 for
H$_{j}\neq$0 in Eq.(\ref{eq9})) To this induced electrical
polarization induced by the magnetic field (linear magnetoelectric
effect), there is another contribution of the induced electrical
polarization which results from the coupling to the magnetic order
parameter (terms in $\frac{-\sigma_{i}}{\kappa_{i}}\eta^{2}$). We
shall discuss these two induced electrical polarization
contributions in section \ref{Discussion}.

\section{Discussion}\label{Discussion}

Using phenomenological Landau analysis, we have shown that for any
magnetically ordered pyroelectric materials (= multiferroics) an
additional electrical polarization should arise upon cooling
trough the magnetic ordering temperature. In previous Landau
theoretical treatment of magnetically ordered ferroelectrics
\cite{Nedlin}, terms linear in the electrical polarization and
quadratic in the magnetic order parameter were not taken into
account. However this kind of terms is allowed for any directions
along which a spontaneous polarization exists. In various
publications terms quadratic in the polarization and in the
magnetic order parameter are considered
\cite{Nedlin,Moskvin,Liu,Hu,Zhong}. These kind of terms are
expected to have a lower contribution (due to their lower degree)
compared to terms like
$\frac{-\sigma_{i}}{\kappa_{i}}P_{i}\eta^{2}$ where $\eta$ is the
magnetic order parameter, $\kappa$ the inverse dielectric
susceptibility and P$_{i}$ the direction of the spontaneous
electrical polarization.

We believe that this kind of term is responsible for the reported
additional electrical polarization for YMnO$_{3}$ and LuMnO$_{3}$
below their respective N\'{e}el temperatures \cite{Park,bas}. Lee
\emph{et al.} explained their results by claiming that they
observed the simultaneous condensation of three order parameters,
spin (S), lattice (L) displacement, and electric dipole (D) moment
\cite{Park}. They claimed that -$\alpha S^{2}L^{2}$ and -L.D terms
condense at the same temperature, with T$_{L}$ = T$_{S}$ for a
sufficient strong spin-lattice interaction strength $\alpha$. If
one considers the literature \cite{Nedlin,Liu,Zhong} and our
results, we see that this statement is incorrect.

The results given by equations (\ref{eq4}) and (\ref{eq10}) are
similar. If we make a series expansion of P$_{x}$ and P$_{y}$ in
terms of $\eta$ to second order, we find analogous expressions for
both expressions. While the magnetic order parameter $\eta$ varies
like $\sqrt{T-T_{N}}$ (mean field approximation), the additional
polarization resulting from the coupling term $\sigma_{i}
P_{i}\eta^{2}$ varies linearly with the temperature (it is
function of $\eta^{2}$; see Eqs. (\ref{eq6})and (\ref{eq10})).
More importantly, this additional polarization arising in the
magnetically ordered phase is a function of the dielectric
susceptibility. Consequently, the higher will be the dielectric
susceptibility of the material under consideration, the higher
will be the additional electrical polarization in the magnetically
ordered phase (see equation (\ref{eq10})). This behavior differs
from the linear magnetoelectric effect since it does not depend on
the magnetic susceptibility \cite{Brown}. In the light of these
considerations, we claim that the observations made for YMnO$_{3}$
and LuMnO$_{3}$ are not isolated cases but should be true for any
magnetically ordered pyroelectric materials.

\section{Conclusion}

In conclusion, we use a phenomenological Landau analysis to
describe the linear magnetoelectric effect in two multiferroic
fluorides, namely KMnFeF$_{6}$ and Ba$_{2}$Ni$_{7}$F$_{18}$. We
use the general expression of the free energy of
Ba$_{2}$Ni$_{7}$F$_{18}$ to discuss in detail the various
contribution to the induced polarization. We show that contrary to
previous reports the biggest contribution of the additional
polarization below the magnetic ordering temperature arises from a
term $\sigma_{i}\eta^{2}$P$_{i}$ where $\eta$ is the magnetic
order parameter and P$_{i}$ the direction of spontaneous
electrical polarization. This additional polarization arising
below the magnetic ordering temperature is characteristic of any
magnetically ordered pyroelectric material. We show that this
induced electrical polarization is proportional to the dielectric
susceptibility of the material. Thus one should see a change in
the slope of the temperature behavior of the electrical
polarization at the magnetic transition with an increase of its
absolute value. We expect that this work will stimulate
experimental investigations of the above reported fluorides but
also of the crystal structure of the magnetoferroelectric
materials below their magnetic ordering temperature.

\section*{Acknowledgements}
We thank T. T. M. Palstra for critical reading of the manuscript.
We thank the anonymous referees for valuable comments which helped
in improving the manuscript.

\end{document}